\documentclass[twocolumn]{revtex4-1}
\usepackage[dvips]{graphicx}
\usepackage{amsmath}
\usepackage{amssymb}
\usepackage{minibox}
\usepackage{microtype}
\usepackage{etoolbox}
\usepackage{hyperref}
\usepackage[usenames,dvipsnames,svgnames,table]{xcolor}
\pdfoutput=1
\hypersetup{urlcolor=blue,citecolor=blue,linkcolor=black,colorlinks=true}

\makeatletter
\apptocmd{\thebibliography}{\global\c@NAT@ctr 10\relax}{}{}
\makeatother

\begin{document}

\title{A one-dimensional liquid of fermions with tunable spin} 

\author{Guido~Pagano$^{1,4}$, Marco~Mancini$^{1,2}$, Giacomo~Cappellini$^1$, Pietro~Lombardi$^{1,2}$, Florian~Sch\"afer$^1$, Hui~Hu$^5$,\\Xia-Ji~Liu$^5$, Jacopo~Catani$^{1,3}$, Carlo~Sias$^{1,3}$, Massimo~Inguscio$^{1,2,3}$, Leonardo~Fallani$^{1,2,3}$}

\affiliation{
\mbox{$^1$LENS European Laboratory for Nonlinear Spectroscopy, 50019 Sesto Fiorentino, Italy}\\
\mbox{$^2$Department of Physics and Astronomy, University of Florence, 50019 Sesto Fiorentino, Italy}\\
\mbox{$^3$INO-CNR Istituto Nazionale di Ottica del CNR, Sezione di Sesto Fiorentino, 50019 Sesto Fiorentino, Italy}\\
\mbox{$^4$Scuola Normale Superiore di Pisa, 56126 Pisa, Italy}\\
\mbox{$^5$Centre for Atom Optics and Ultrafast Spectroscopy, Swinburne University of Technology, Melbourne 3122, Australia}
}

\begin{abstract}
Published in Nature Physics {\bf 10}, 198 (2014), \href{http://www.nature.com/doifinder/10.1038/nphys2878}{http://www.nature.com/doifinder/10.1038/nphys2878}.
\end{abstract}

\maketitle

{\bf 
Correlations in systems with spin degree of freedom are at the heart of fundamental phenomena, ranging from magnetism to superconductivity. The effects of correlations depend strongly on dimensionality, a striking example being one-dimensional (1D) electronic systems, extensively studied theoretically over the past fifty years \cite{giamarchi2004,yang1967,gaudin1967,sutherland1968,luttingerreview,fiete2007,imambekov2012}. However, the experimental investigation of the role of spin multiplicity in 1D fermions -- and especially for more than two spin components -- is still lacking. Here we report on the realization of 1D, strongly-correlated liquids of ultracold fermions interacting repulsively within SU($N$) symmetry, with a tunable number $N$ of spin components. We observe that static and dynamic properties of the system deviate from those of ideal fermions and, for $N>2$, from those of a spin-1/2 Luttinger liquid. In the large-$N$ limit, the system exhibits properties of a bosonic spinless liquid. Our results provide a testing ground for many-body theories and may lead to the observation of fundamental 1D effects \cite{recati2003}.
}

\begin{figure}[b!]
\begin{center}
\includegraphics[width=\columnwidth]{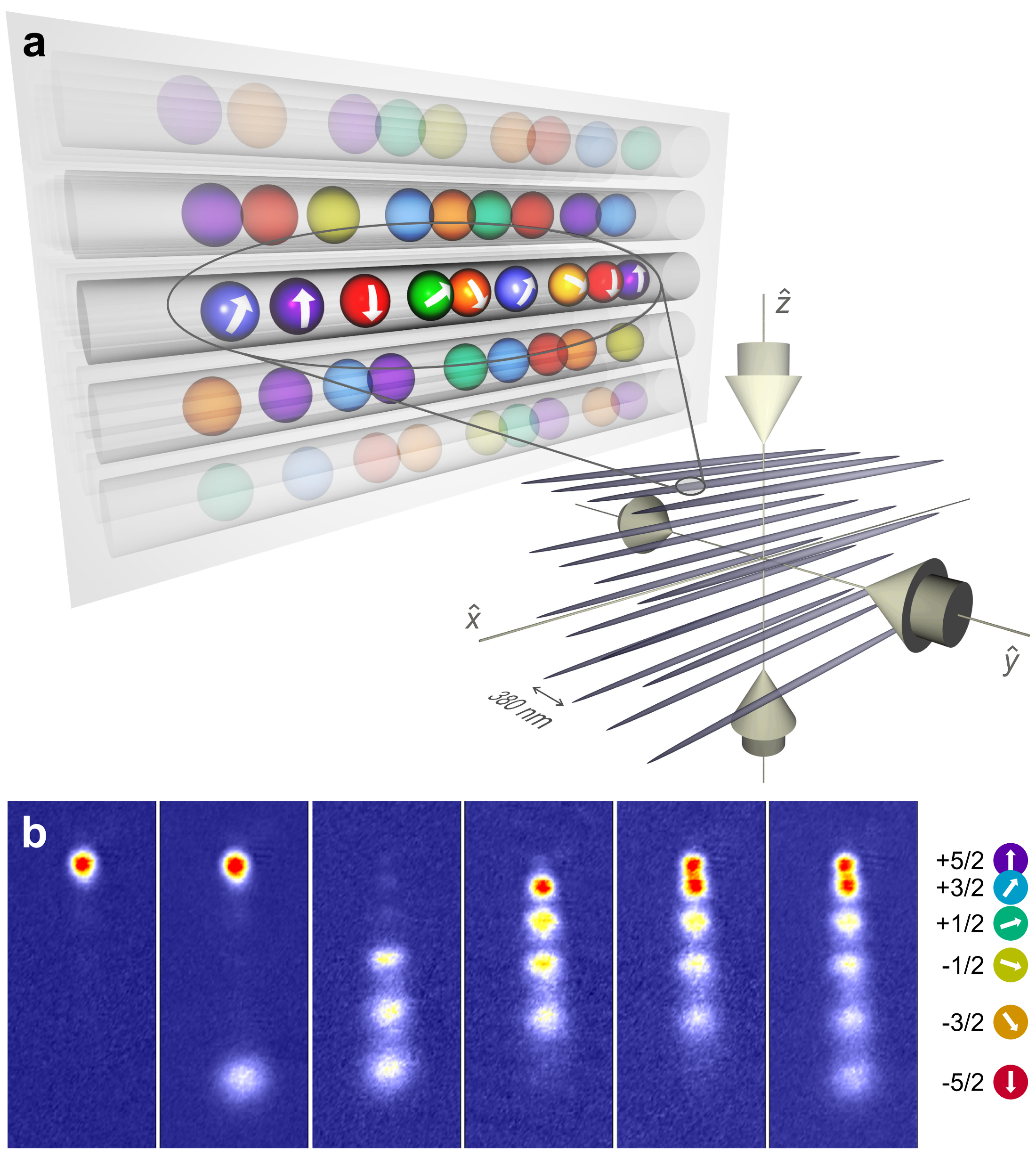}
\end{center}
\caption{{\bf Ultracold 1D fermionic liquids with tunable spin.} {\textbf a}, A 2D optical lattice is used to create an array of independent quantum wires of ultracold $^{173}$Yb with 6 possible nuclear spin orientations. {\textbf b}, The nuclear spin of the atoms can be manipulated with optical pumping techniques, resulting in a tunable number of spin components, and analyzed with optical Stern-Gerlach detection (see Supplementary Information).} \label{fig:scheme}
\end{figure}

One-dimensional quantum systems show specific, sometimes counterintuitive behaviours that are absent in the three-dimensional world. These behaviours, predicted by many-body models of interacting bosons \cite{lieb1963} and fermions \cite{yang1967,gaudin1967,sutherland1968}, include the ``fermionization'' of bosons \cite{girardeau1960} and the separation of spin and density (most commonly referred to as ``charge'') branches in the excitation spectrum of interacting fermions. The last phenomenon is predicted within the celebrated Luttinger liquid model \cite{luttingerreview}, which describes the low-energy excitations of interacting spin-1/2 fermions. Although the Luttinger approach describes qualitatively the physics of a number of 1D systems \cite{mceuen1999,west1996}, the problem of how to extend it to a more detailed description of real systems has puzzled physicists over the years \cite{imambekov2012}. In this exploration the physics of spin has played a key role.

Ultracold atoms have proved to be a precious resource to study 1D physics, as they afford exceptional control over experimental parameters. Most of the experiments so far have been performed with spinless bosons, which for instance led to the realization of a Tonks-Girardeau gas \cite{weiss2004,paredes2004}. On the other hand, 1D ultracold fermions are a promising system to observe a number of elusive phenomena, like Stoner's itinerant ferromagnetism \cite{ho2013} and the physics of spin-incoherent Luttinger liquids \cite{fiete2007}. However, only a few pioneering works, dealing with spin-1/2 particles \cite{moritz2005,hulet2010,jochim2012}, have been reported so far.

In parallel, ultracold two-electron atoms have been recently proposed for the realization of large-spin systems with SU($N$) interaction symmetry \cite{cazalilla2009,gorshkov2010}, and first experimental investigations have been reported \cite{taie2012}. This novel platform enables the simulation of 1D systems with high degree of complexity, including spin-orbit-coupled materials \cite{troyer1999} or SU($N$) Heisenberg and Hubbard chains \cite{bonnes2012,messio2012}. Moreover, the investigation of these multi-component fermions is relevant for the simulation of field theories with extended SU($N$) symmetries \cite{banerjee2013}.

In this Letter we report on the realization of 1D quantum wires of repulsive fermions with a tunable number of spin components, which are created by tightly trapping ultracold $^{173}$Yb atoms in a two-dimensional optical lattice (Fig. \ref{fig:scheme}a). The purely nuclear spin $I=5/2$ of $^{173}$Yb results both in the independence of the interaction strength from the nuclear spin state and in the absence of spin-changing collisions. The latter feature is particularly important for our experiments, since it implies the stability of any spin mixture. The atoms experience an axial harmonic confinement with (angular) frequency $\omega_x  \approx 2\pi \times 80$ Hz and a strong radial confinement with $\omega_\perp=2\pi \times 25$ kHz, resulting in the occupation of the radial ground state. We use optical spin manipulation and detection techniques (see Supplementary Information) to prepare the system in an arbitrary number $N \le 2I+1=6$ of spin components (Fig. \ref{fig:scheme}b), thus realizing different SU($N$) symmetries. We directly compare systems with different $N$, keeping the atom number per spin component $N_\mathrm{at} \simeq 6000$ ($\approx 20$ atoms per spin component in the central wire) and all the other parameters constant. This approach enables us to evidence how the physics of a strongly-interacting 1D fermionic system changes as a function of $N$.

\paragraph*{\bf Momentum distribution.}

\begin{figure}[t!]
\begin{center}
\includegraphics[width=0.97\columnwidth]{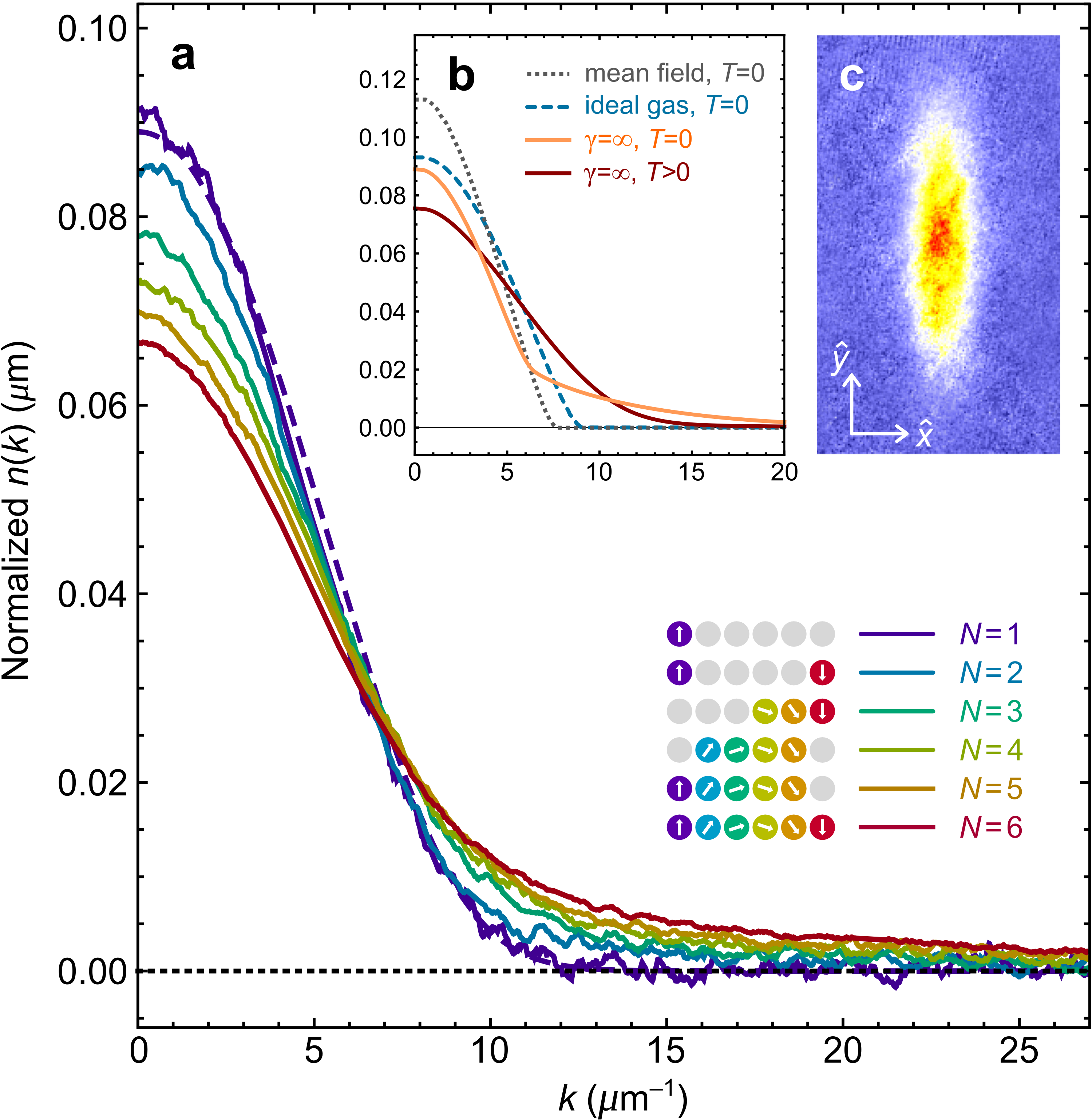}
\end{center}
\caption{{\bf Momentum distribution of the 1D fermions.} {\textbf a}, Solid lines: momentum distribution $n(k)$ measured with time-of-flight absorption imaging for different $N$ and the same atom number $N_\mathrm{at}$ per spin component (each curve results from the average of 30-50 experimental images, after integration along the $\hat{y}$ axis and normalization to unity area). Dashed line: theoretical curve for $N=1$ based on the ideal trapped Fermi gas theory, after averaging over the inhomogeneous distribution of atoms in the different wires. {\textbf b}, Theoretical $n(k)$ for the $N=2$ system derived from different models (see Supplementary Information): ideal Fermi gas at $T=0$ (dashed), mean-field treatment of finite interactions at $T=0$ (dotted), full many-body problem for infinite repulsion both for $T=0$ (light solid, from \cite{ogata1990}) and $T_S \ll T \ll T_F$ (dark solid, from \cite{cheianov2005}). While the mean-field curve shows an opposite behaviour from the one observed in the experiment, the many-body curves account for the observed broadening. {\textbf c}, Averaged absorption image (the $\hat{x}$ axis denotes the direction of the wires).} \label{fig:nk}
\end{figure}

We investigate the correlations in the 1D wires by observing the momentum distribution $n(k)$ ($k$ is the momentum divided by the reduced Planck's constant $\hbar$). We measure this quantity by extinguishing the trapping light and imaging the atomic cloud after a ballistic expansion, as done in previous works to measure $n(k)$ of a Tonks-Girardeu gas \cite{paredes2004}. A typical image is reported in Fig. \ref{fig:nk}c, where $\hat{x}$ denotes the wire axis. Integration over $\hat{y}$ results in the $n(k)$ curves plotted in Fig. \ref{fig:nk}a for different $N$ (the curves are normalized to have the same unity area). In the noninteracting case $N=1$ the data (solid blue) are very well accounted for by the theory of a trapped ideal Fermi gas (dashed blue, see Supplementary Information). Increasing $N$, we observe a clear monothonic broadening of $n(k)$, with a reduction of the weight at low $k$ and a slower decay of the large-$k$ tails. 

The observed $n(k)$ broadening arises from a pure effect of correlations that goes beyond standard mean-field physics. In order to give a qualitative understanding of this phenomenon, we consider spin-1/2 fermions with infinite repulsion. In this limit, the density-density correlation function $G_{\uparrow \downarrow}(d)=\langle \hat{n}_\uparrow(x+d)\;\hat{n}_\downarrow(x) \rangle$ (where $\hat{n}_\uparrow(x)$ and $\hat{n}_\downarrow(x)$ are the density operators for the two spin components) falls to zero for $d \rightarrow 0$ as $G_{\uparrow \uparrow}(d)$ does in the case of a spin-polarized gas, thus mimicking the effects of a Pauli repulsion between distinguishable particles. This ``fermionization'', restricting the effective space which is available to the particles, causes them to populate states with {\it larger} momentum \cite{ogata1990,cheianov2005}. We note that an opposite behaviour would be predicted by a mean-field treatment of interactions neglecting correlations between trapped fermions: the effectively weaker confinement along $\hat{x}$ induced by the atom-atom repulsion would lead to more extended single-particle wavefunctions, hence to a {\it decreased} width of $n(k)$ (see Fig. \ref{fig:nk}b). For $N=2$ the interaction regime of our 1D samples is described by the parameters $\gamma \simeq 4.8$ and $K \simeq 0.73$ (see Supplementary Information), lying in the strongly-correlated regime between the ideal Fermi gas ($\gamma=0$, $K=1$) and a fully fermionized gas ($\gamma=\infty$, $K=0.5$). 

The details of $n(k)$ depend nontrivially on the temperature, owing to the thermal population of spin excitations. The temperature regime for our experiments, $T \simeq 0.3\, T_F$ (where $T_F$ is the Fermi temperature), is slightly below the predicted temperature scale $T_S \simeq 0.4\, T_F$ for spin excitations (see Supplementary Information), in the crossover between the spin-ordered regime for $T \ll T_S$ and that of a spin-incoherent Luttinger liquid for $T \gg T_S$ \cite{fiete2007}. Fig. \ref{fig:nk}b shows the theoretical $n(k)$ for $N=2$ and infinite repulsion in the limiting regimes $T=0$ and $T \gg T_S$ (light and dark solid curves, derived from \cite{ogata1990} and \cite{cheianov2005}, respectively). While both curves show an evident $n(k)$ broadening, in accordance with our observations, their shape is different and can be explained in terms of a modified {\it effective} Fermi momentum \cite{feiguin2010}. Exact calculations for finite interactions and finite temperatures are challenging, thus making our system a profitable quantum simulation resource for the fundamental problem of 1D interacting fermions.

\paragraph*{\bf Probing excitations.}

\begin{figure}[t!]
\begin{center}
\includegraphics[width=0.97\columnwidth]{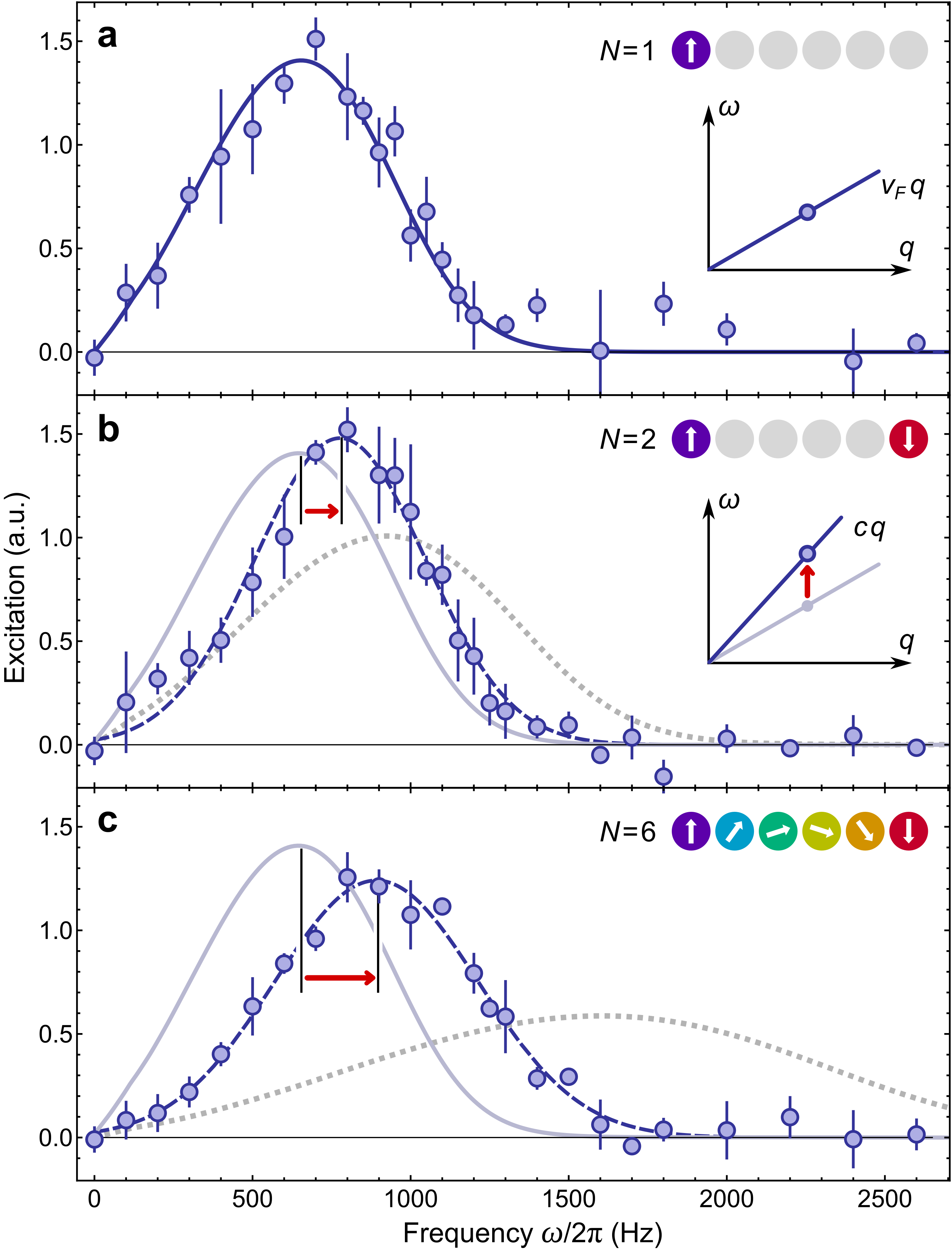}
\end{center}
\caption{{\bf Excitation spectra of the 1D fermions.} The points show the measured increase in atomic momentum after a Bragg excitation with energy $\hbar \omega$ and momentum $\hbar q \simeq 0.2 \, \hbar k_F^0$ (see text) for $N=1$ ({\bf a}), $N=2$ ({\bf b}) and $N=6$ ({\bf c}) spin components and same atom number $N_{\mathrm{at}}$ per spin component. The error bars are standard deviations over up to 5 repeated measurements per frequency. The solid line is the calculated response function for the ideal Fermi gas $N=1$, while the dotted lines show the calculation in the limit of infinite repulsion. The dashed lines are Gaussian fits to the experimental points, in order to guide the eye and to extract the peak excitation frequency. Both the experimental and theoretical spectra have been normalized to unity area. The graphs in the inset show a sketch of the excitation spectrum at low $q$ for the ideal Fermi gas ({\bf a}) and for the two-component Luttinger liquid ({\bf b}) with repulsive interactions. The red arrows indicate the shift in the excitation resonance.} \label{fig:spectra}
\end{figure}

A distinctive feature of 1D fermions is the existence of a well-resolved excitation spectrum at small momentum transfer $\hbar q \ll \hbar k_F$ (where $k_F$ is the Fermi wavevector). Number-conserving excitations in the ideal 1D Fermi gas correspond to  particle-hole pairs with energy $\hbar \omega = v_F \hbar q$, where $v_F=\hbar k_F /m$ is the Fermi velocity (see inset of Fig. \ref{fig:spectra}a). This physical picture changes in the case of an interacting spin mixture, since excitations acquire a purely collective nature. According to the Luttinger theory, the spectrum of phononic excitations is still described by a linear dispersion $\omega = c q$, where $c=v_F/K$ is a renormalized sound velocity \cite{giamarchi2004}. In a two-component Luttinger liquid with contact repulsion one has $0.5 < K  < 1 $. This yields a sound velocity that is larger than $v_F$ (see inset of Fig. \ref{fig:spectra}b), corresponding to an increased stiffness of the many-body state.

We have characterized the excitations of the fermionic wires by performing Bragg spectroscopy. This technique, relying on inelastic light scattering, allows the selective excitation of density waves with energy $\hbar \omega$ and momentum $\hbar q$ (see Supplementary Information). Fig. \ref{fig:spectra}a shows the measured spectrum for $N=1$ at low momentum transfer $\hbar q \simeq 0.2 \, \hbar k_F^0$ (being $k_F^0$ the peak Fermi wavevector in the central wire). A clear resonance is observed, in excellent agreement with the calculated response for ideal fermions (solid line, with no free parameters). For $N=2$ the resonance is clearly shifted towards higher frequencies (Fig. \ref{fig:spectra}b), as expected from the Luttinger theory. The measured shift $(+15 \pm 4)\%$ agrees with the expected $(+10 \pm 2)\%$ shift in the sound velocity predicted on the basis of the Luttinger theory for a trapped system (see Supplementary Information). For $N=6$ the spectrum shows a much larger shift $(+33 \pm 4)\%$ (Fig. \ref{fig:spectra}c), which disagrees with the predictions for $N=2$, signalling an increased effect of interactions, in qualitative accordance with the $n(k)$ change of Fig. \ref{fig:nk}. We also plot the calculated spectra for trapped fermions with infinite interactions (dotted lines in Fig. \ref{fig:spectra}b-c), which evidence how the measured spectra lie between the response of the ideal Fermi gas and that of a fermionized system.

\paragraph*{\bf Collective mode frequencies.}

\begin{figure}[t!]
\begin{center}
\includegraphics[width=\columnwidth]{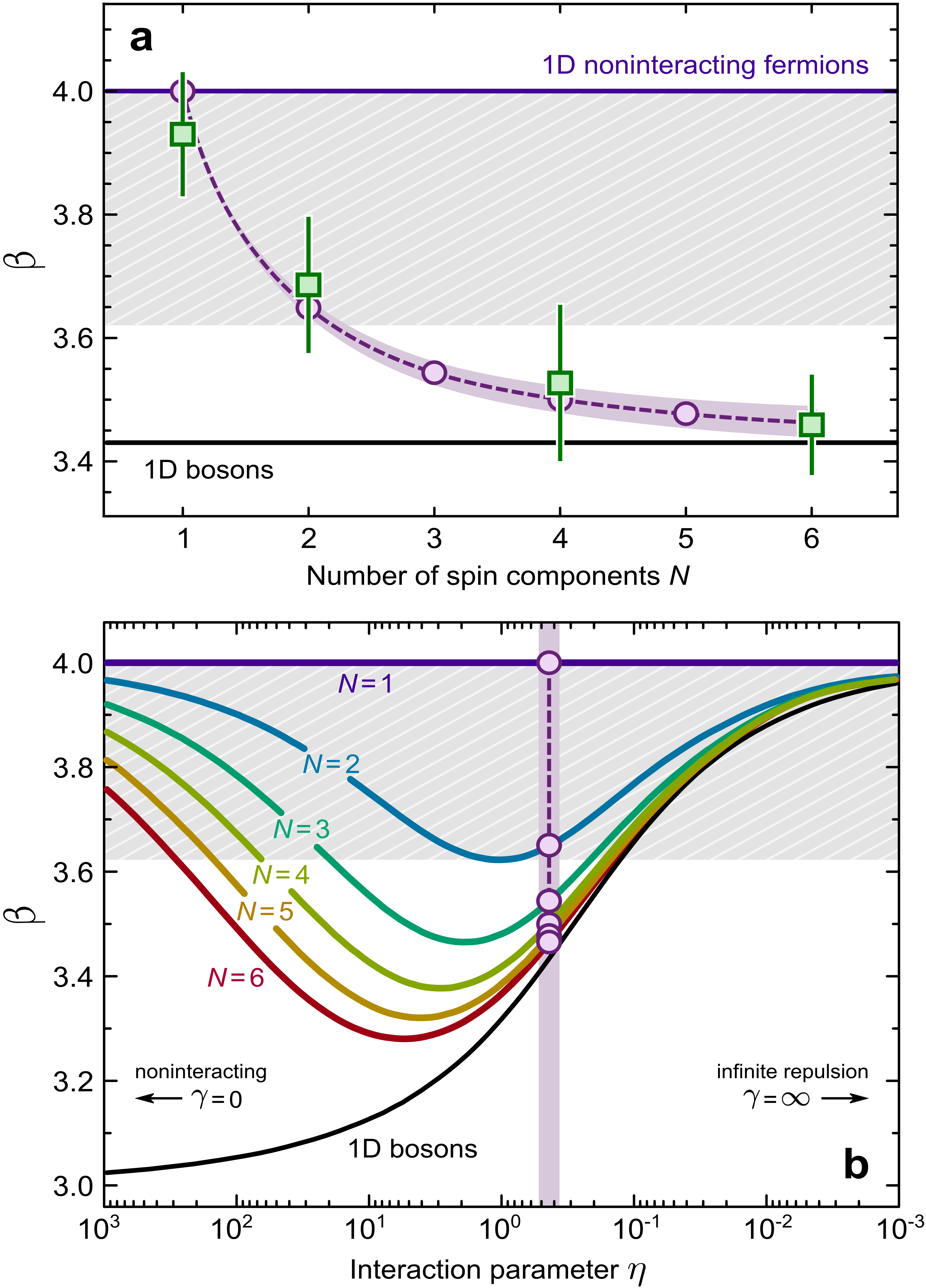}
\end{center}
\caption{
{\bf Breathing oscillations.} The quantity which is plotted in the graphs is the squared ratio $\beta=\left(\omega_B/\omega_x\right)^2$ of the breathing frequency $\omega_B$ to the trap frequency $\omega_x$. {\bf a}, The squares show the experimental data, as a function of $N$, obtained as the weighted mean over sets of up to 9 repeated measurements (the error bars indicate the standard deviation of the weighted mean). The circles show the theoretical predictions for the average interaction parameter (defined in the text) $\eta=0.44$ of our experiment. The dashed line is a guide to the eye, while the height of the violet shaded region indicates the uncertainty on the theoretical values resulting from the experimental uncertainty $\Delta \eta=0.08$ (coming from the measured atom number and trapping frequencies). The upper horizontal line shows the theoretical value for the noninteracting Fermi gas ($N=1$), while the lower line shows the result for 1D spinless bosons. {\bf b}, The lines show the theoretical dependence of $\beta$ on the interaction parameter $\eta$ (large $\eta$ values correspond to small $\gamma$, i.e. a small effect of interactions). The circles show the predicted values for our average interaction parameter (also shown in panel {\bf a}), while the width of the violet shaded region indicates the experimental uncertainty $\Delta \eta=0.08$. In both panels the height of the grey region shows the range of $\beta$ for $N=2$ and {\it any} possible value of the repulsion strength.} \label{fig:frequencies}
\end{figure}

More insight into the physics of multicomponent 1D fermions can be gained by studying low-energy breathing oscillations in which the cloud radius oscillates in time. We measure the frequency of this collective mode by suddenly changing the trap frequency and measuring the time evolution of the radius. In Fig. \ref{fig:frequencies}a we plot the measured squared ratio $\beta=\left(\omega_B/\omega_x \right)^2$ of the breathing frequency $\omega_B$ to the trap frequency $\omega_x$ as a function of $N$ (squares). For $N=1$ the measured value is in good agreement with the expected value $\beta=4$ for ideal fermions (upper horizontal line). Increasing $N$ our data clearly show a monothonic decrease of $\beta$, induced by the repulsive interactions in the spin mixture.

The dependence of $\beta$ on the interaction strength is remarkably nontrivial, already for $N=2$, as first predicted in Ref. \cite{astrakharchik2004}. Indeed, $\beta=4$ in both the limiting cases of an ideal gas ($\gamma=0$) and a fermionized ($\gamma=\infty$) system, while for finite repulsion it is expected to exhibit a nonmonotonic behaviour with a minimum at finite interaction strength. The theoretical curves in Fig. \ref{fig:frequencies}b show the expected dependence of $\beta$ on the interaction parameter $\eta=N^1_\mathrm{at}\left( a_\mathrm{1D}/a_x \right)^2$ (where $N^1_\mathrm{at}$ is the number of atoms per wire,  $a_\mathrm{1D}$ is the 1D scattering length and $a_x$ is the trap oscillator length). We have derived these results by combining a Bethe Ansatz approach with the exact solution of the hydrodynamic equations describing a 1D fermionic liquid with $N$ components (see Supplementary Information). As $N$ is increased, the curves exhibit an increasingly larger redshift of $\beta$, and for $N \rightarrow \infty$ they asymptotically approach the curve for 1D spinless bosons. The circles indicate the theoretical values for the average $\eta=0.44$ in our experiment. The agreement between experiment and theory is excellent, as shown in Fig. \ref{fig:frequencies}a (we note that for $N=2$ both theory and experiment agree with the results of Ref. \cite{astrakharchik2004}).

The experimental data, accompained by our theoretical curves, clearly show that changing $N$ causes quite different effects from those induced by simply changing the interaction strength in a $N=2$ mixture. In fact, by increasing $N$, the constraints of the Pauli principle become less stringent and the number of binary-collisional partners increases, causing the system to acquire a more ``bosonic'' behaviour. Our experimental value at $N=6$ clearly falls out of the range of $\beta$ expected for a $N=2$ liquid (grey regions in Fig. \ref{fig:frequencies}) and already approaches the value expected for 1D spinless bosons. This bosonic limit for $N \rightarrow \infty$ is a remarkable property of multi-component 1D fermions that has been pointed out theoretically only very recently \cite{yang2011} and that our experimental system is capable to clearly evidence.

\paragraph*{\bf Concluding remarks.}

The possibility of tuning the number of spin components allows us to study different regimes of interplay between Fermi statistics and degree of distinguishability in this novel 1D system. In a quantum simulation perspective, this realization provides a powerful testbench for large-spin models and opens to the investigation of fundamental effects, ranging from spin dynamics to novel magnetic phases.

\appendix*

\section*{Acknowledgements}

We would like to thank all the members of the Ultracold Quantum Gases group in Florence for fruitful discussions. We are indebted to P. Calabrese for enlightening discussions on the physics of one-dimensional quantum systems, and to P. Cancio, G. Giusfredi and P. De Natale for early valuable contributions to the experimental setup. We also thank M. Dalmonte, A. Recati and M. Polini for valuable discussions. This work has been
financially supported by IIT Seed Project ENCORE, ERC Advanced Grant DISQUA, EU FP7 Integrated Projects AQUTE and SIQS, MIUR Project PRIN2009, ARC Discovery Projects (Grants No. DP0984522 and No. DP0984637) and NFRP-China (Grant No. 2011CB921502).

\section*{Author Contributions}
All authors contributed to the writing of the manuscript. G.P., M.M., G.C., P.L., F.S., J.C., C.S., M.I. and L.F. built the experimental setup, performed the measurements and analysed the data. H.H. and X.-J.L. carried out the theoretical derivation of the breathing frequencies.

\section*{Additional Information}
Correspondence and requests for materials should be addressed to L.F. (fallani$@$lens.unifi.it).

\section*{Competing financial interests}
The authors declare no competing financial interests. \phantom{The authors declare no competing financial interests. }

\renewcommand{\thefigure}{\arabic{figure}}
 \setcounter{figure}{0}
\renewcommand{\theequation}{\arabic{equation}}
 \setcounter{equation}{0}
 \renewcommand{\thesection}{\Roman{section}}
\setcounter{section}{0}
\renewcommand{\thetable}{\arabic{table}}
 \setcounter{table}{0}

\onecolumngrid

\newpage


\begin{center}
{\bf \large Supplementary Information for\\
`A one-dimensional liquid of fermions with tunable spin'}

\bigskip
Guido Pagano, Marco Mancini, Giacomo Cappellini, Pietro Lombardi, Florian Sch\"afer, Hui Hu\\
Xia-Ji Liu, Jacopo Catani, Carlo Sias, Massimo Inguscio, Leonardo Fallani
\end{center}

\bigskip
\bigskip
\twocolumngrid

\noindent
{\bf Preparation of multi-component one-dimensional $^{173}$Yb fermions.}
The $^{173}$Yb ground state is characterized by a purely nuclear spin $I=5/2$ and, similarly to the other alkaline-earth fermionic isotopes, by the absence of hyperfine interaction, which determines the independence of the scattering length on the nuclear spin orientation. This also results in the collisional stability of any nuclear spin mixture, caused by the absence of spin-changing collisions. Optical manipulation schemes, involving multi-step optical pumping before evaporative cooling and selective removal of unwanted spin states at the end of evaporation, allow us to initialize the system in a well-defined number of spin components with equal population. The nuclear spin composition is analyzed by means of optical Stern-Gerlach detection (see Fig. \ref{fig:scheme}b) exploiting the spin-selective optical dipole force of a focused near-resonant laser beam \cite{taie2010}. The beam is circularly-polarized, with a detuning of --560 MHz from the $^1 S_0$ $\rightarrow$ $^3P_1$ ($F'=7/2$) transition, 50 $\mu$m waist and 4 mW power. The difference in (false-colour) intensity of the density peaks in Fig. \ref{fig:scheme}b is caused by the different shape of the cloud after Stern-Gerlach deflection (caused by the non-uniform intensity gradient): integrating the density profiles to obtain the total number of atoms gives the same population for each of the spin states within an experimental uncertainty $<5\%$. 

The starting point for our experiment is an ultracold $^{173}$Yb trapped gas at $T/T_F < 0.3$ (where $T_F$ is the Fermi temperature) and $N_{\mathrm{at}}\simeq 6000$ atoms per spin component. By slowly increasing the intensity of a 2D optical lattice in $\delta t = 150$ ms we produce an array of $\approx 600$ independent 1D fermionic wires, weakly confined by a harmonic potential along the wire axis. The final depth of the optical lattice (operating at a wavelength $\lambda=759$ nm) is $V_0=40E_R$ (where $E_R=h^2/2m\lambda^2$ is the recoil energy), resulting in a tight radial confinement with (angular) trapping frequency $\omega_\perp =  2\pi \times 25$ kHz. Since the energy $\hbar \omega_\perp $ is larger than the Fermi energy by more than one order of magnitude, the occupation of excited radial modes is negligible, which makes our wires truly one-dimensional. Furthermore, the tunnelling time between different wires is on the order of several seconds, much longer than the timescale of the experiments, which makes the wires effectively decoupled from each other.

\vspace{4mm}
\noindent
{\bf Interaction regime for two-component fermions.}
The interaction regime for a two-component 1D Fermi gas with contact interactions can be described by the dimensionless  parameter
\begin{equation}
\gamma = \frac{1}{|a_{\mathrm{1D}}| n} \; ,
\end{equation}
where $n$ is the 1D density (per spin component) and $a_{\mathrm{1D}}$ is the 1D scattering length \cite{olshanii1998}, given by
\begin{equation}
a_{\mathrm{1D}} = -\frac{a_\perp^2}{a} \left( 1 - 1.0326 \frac{a}{a_\perp} \right) \; ,
\end{equation}
where $a$ is the 3D scattering length (+10.58 nm for $^{173}$Yb) and $a_\perp=\sqrt{\hbar /m \omega_\perp}$ is the radial harmonic oscillator length. In our system we calculate a mean interaction parameter $\gamma \simeq 4.8$ (averaged over the atomic density, see next section), which lies in the strongly-interacting regime between the ideal Fermi gas ($\gamma=0$) and the fully-fermionized gas ($\gamma=\infty$).  It is interesting to note that in 1D the effect of collisions between particles is increased as the density is decreased, contrarily to the behaviour in higher dimensions. As a consequence, the Fermi pressure, which limits the density $n$ of a trapped fermionic gas by keeping its constituent particles away from each other, in 1D has the counterintuitive effect of making a system of fermions more interacting than one of bosons (and of a classical gas as well). 

Another relevant quantity is the Luttinger parameter $K$, which enters the low-energy Hamiltonian. After bosonization \cite{giamarchi2004}, the Hamiltonian in the charge (density) sector reads
\begin{equation}
H = \int dx \left[ \left( \frac{\pi c K}{2}\right) \Pi(x)^2 + \left( \frac{c}{2 \pi K}\right) \left( \partial_x \phi(x)\right)^2 \right] \; ,
\label{eq:luttinger}
\end{equation}
which describes the free excitations of the Luttinger liquid, formally equivalent to the modes of vibration of an elastic string ($\phi(x)$ and $\Pi(x)$ are conjugate bosonic fields describing the collective density excitations of the liquid). We use the word ``charge'' for a system of neutral atoms, in analogy to the common use in condensed-matter physics, to indicate excitations involving a perturbation of the total fermionic density (as opposed to excitations in the spin sector, where the total density is constant and the local magnetization is perturbed). The $K$ parameter determines the power-law divergencies in some relevant properties of the Luttinger liquid (for instance, the momentum distribution $n(k)$ around the Fermi momentum) and characterizes the sound velocity of its collective (charge) excitations, given by $c=v_F/K$. In our system $K \simeq 0.73$ \cite{recati2003b}, which interpolates between the behaviour of the ideal Fermi gas ($K=1$) and that of a system with infinitely-strong contact repulsion ($K=0.5$).

\vspace{4mm}
\noindent
{\bf Effect of inhomogeneity and finite temperature.}
The presence of an axial trapping potential, with (angular) frequency $\omega_x$ ranging from $2\pi\times 60$ Hz to $2\pi\times 100$ Hz (the precise number depending on the particular experiment), causes the 1D fermionic wires to have a nonuniform density profile $n(x)$. Additionally, moving out of the trap center, the number of atoms per wire is decreasing from a maximum of $\approx 20$ (per spin component) in the central wire, to a vanishing occupation of the more peripheral wires. In order to interpret our experimental findings, we have considered these two effects very carefully.

The number of fermions $N_{ij}$ in each wire has been calculated in the noninteracting case by determining the lowest-energy configuration of $N_{ij}$ which satisfies both Fermi statistics and the constraint on the total atom number $N_{at}=\sum_{i,j} N_{ij}$. The calculation has been performed by taking into account a wire-dependent energy potential offset $\epsilon_{i,j}=\frac{1}{2}m \omega_y^2 d^2 i^2 + \frac{1}{2}m \omega_z^2 d^2 j^2$, where $m$ is the particle mass, $d=\lambda/2$ is the lattice spacing and $\omega_{y,z}$ are the (angular) frequencies of the slowly-varying harmonic trapping potentials in the orthogonal directions to the wires. 

The resulting atom distribution $N_{ij}$ is a very good description of the experimental system, as we can verify by looking at the agreement between the experimental points and the theoretical curves for $N=1$ in Figs. \ref{fig:nk}a and \ref{fig:spectra}a, where the only adjusted parameter is the temperature (see below). Those curves have been calculated by taking the momentum distribution $n(k)$ and the dynamic structure factor for a homogeneous ideal gas $S(q,\omega)$, respectively, and treating the effects of the axial harmonic potential in a local density approximation (LDA), i.e. assuming that locally the properties of the system are described in terms of a local Fermi wavevector $k_F(x) =\pi n(x)$. Then we have averaged the result over all the wires, taking the number of atoms per wire as weight. The temperature assumed for both the calculations is $T=0.3\,T_F$, which well agrees with the $T/T_F$ ratio measured in the 3D Fermi gas after ramping down the lattices from a fit of a polylogarithmic function \cite{demarco2001} to the time-of-flight density distribution.

\vspace{4mm}
\noindent
{\bf Momentum distribution of the 1D interacting liquids.}
The momentum distribution of the 1D interacting liquids is measured by suddenly switching off the optical trapping potentials and detecting the atomic density with absorption imaging after a time-of-flight $t_{\mathrm{TOF}}=23$ ms of ballistic expansion. The expansion is in the far-field regime, which maps the initial momentum distribution $n(k)$ to the expanded density in coordinate space. We neglect the initial size of the trapped sample since it contributes by only $\sim \left( \omega t_{\mathrm{TOF}} \right)^{-2}/2 \simeq 0.3 \%$ to the size of the expanded cloud. 

We rule out a possible explanation of the observed changes in $n(k)$ in terms of different temperatures for different $N$. Indeed, we have verified that, after slowly ramping down the lattices in $\delta t = 150$ ms to recover a 3D Fermi gas, the temperature measured for the different spin mixtures has the same value $T=0.3\,T_F$ for all $N$ within the experimental uncertainties. In this 3D regime the effects of interactions are very weak. As a matter of fact, for 3D Fermi gases we have not detected any significative change in $n(k)$ as a function of $N$. This observation both makes the temperature measurement in 3D reliable and demonstrates that the observed increase in width comes from the increased correlations in the interacting 1D systems.

The theoretical $n(k)$ curves for $N=2$ in Fig. \ref{fig:nk}b have been calculated by taking into account the increased width of the trapped interacting spin mixture following the results of \cite{astrakharchik2004}. This effect leads to a decrease of the density (with respect to the noninteracting case) and to a reduction of the Fermi wavevector $k_F$, which influences both the momentum distribution and the excitation spectrum. The mean-field $n(k)$ curve has been calculated from the momentum distribution of an ideal Fermi gas in an effectively weaker harmonic trap, which accounts for the increase in spatial width calculated for our interactions strength. The $\gamma=\infty$ curves have been calculated by taking the results of Refs. \cite{ogata1990} and \cite{cheianov2005} for a homogeneous system and treating them with LDA + average over the atom distribution in the wires.

The temperature of the 1D fermionic wires is slightly below the characteristic temperature scale $T_S$ for spin excitations, over which the spin-incoherent Luttinger liquid regime begins \cite{cheianov2005}. This spin temperature corresponds to the maximum energy difference between different spin configurations. In the limit of large $\gamma$ the bandwidth of the spin excitations can be estimated \cite{cheianov2005} as $\epsilon_S = (8 \ln 2 ) k_B T_F / 3 \gamma$, which corresponds to a spin temperature $T_S = \epsilon_S/k_B \simeq 0.4 \, T_F$ for our experimental parameters.

\vspace{4mm}
\noindent
{\bf Bragg spectroscopy and excitation spectra.}
Bragg spectroscopy is performed by exciting the atomic cloud with two off-resonant laser beams \cite{stenger1999,steinhauer2002,clement2009}. This technique allows the selective excitation of density waves with energy $\hbar \omega = \hbar(\omega_1-\omega_2)$ and momentum $\hbar \mathbf{q} = \hbar (\mathbf{k}_1-\mathbf{k}_2)$, where $\omega_i$ and $\mathbf{k}_i$ are the (angular) frequencies and the wavevectors of the two Bragg beams, respectively. In order to access the low-momentum part of the spectrum, the $\lambda=759$ nm Bragg beams are aligned at a small angle, resulting in a momentum transfer $\hbar q \simeq 0.2 \, \hbar k_F^0$ along the wire axis (being $k_F^0$ the Fermi wavevector corresponding to the peak density of the central wire). Bragg pulse length and intensity are chosen in such a way to result in a combined interaction-time and power broadening $<100$ Hz, much less than the width of the measured spectra. In order to both minimize the effect of laser phase fluctuations and increase the signal, the spectrum of each of the two Bragg beams contains both the frequencies $\omega_1$ and $\omega_2$, resulting in both left-moving and right-moving excitations. We quantify the Bragg excitation by measuring the momentum transferred to the cloud by time-of-flight imaging. This quantity is proportional to the imaginary part of the response function $\chi(q,\omega)$, which is directy related to the dynamic structure factor $S(q,\omega)$ by the relation
\begin{equation}
\mathrm{Im}\left[ \chi(q,\omega)\right]\propto \left[ S(q,\omega)-S(-q,-\omega)\right] \; , 
\label{eq:responsebragg}
\end{equation}
which reduces to $S(q,\omega)$ in the small-temperature limit \cite{brunello2001}. We note that, since the Bragg light is far detuned with respect to any atomic resonance, the Bragg perturbation acts equally on the different nuclear spin states, i.e. we are only exciting {\it charge} modes (density waves) and not {\it spin} modes (spin waves) \cite{hoinka2012}, which would propagate with a different velocity.
 
Experimentally, we characterize the shift in the sound velocity by determining the frequency $\omega_{\mathrm{peak}}$ corresponding to the excitation peak. This quantity is obtained by performing a Gaussian fit to the experimental points (blue dashed lines in Figs. \ref{fig:spectra}b,c). By considering the full dynamic structure factor in the noninteracting case, we have theoretically verified that $\omega_{\mathrm{peak}}$ depends linearly on the excitation momentum $\hbar q$ up to the value used in our measurements. This means that, despite the average over different inhomogeneous systems and the finiteness of our momentum transfer, we are effectively probing the low-$q$ regime of linear dispersion in the excitation spectrum. 

The theoretical shift in the sound velocity expected for the low-$q$ excitation spectrum in the $N=2$ case has been evaluated following the results of \cite{astrakharchik2004,recati2003b} for the homogeneous system and performing the same LDA + averaging procedure described previously. The curves for infinite repulsion in Figs. \ref{fig:spectra}b,c have been derived following the fermionization picture, i.e. by calculating the quantity in Eq. (\ref{eq:responsebragg}) for the dynamic structure factor $S(q,\omega)$ of an ideal Fermi gas with $N \times N_{at}$ particles.

\vspace{4mm}
\noindent
{\bf Theoretical derivation of the breathing frequencies.}
The low-energy dynamics of density fluctuations of a 1D fermionic wire is described by the Luttinger liquid model Eq. (\ref{eq:luttinger}). The gradient of the phase field $\phi \left( x,t\right) $ gives rise to the density fluctuation, $\delta n\left( x,t\right) =-\partial _x\phi \left( x,t\right) $, while the momentum operator $\Pi \left( x,t\right) $ is proportional to the density current, $j\left( x,t\right) =c(x)K(x)\Pi \left( x,t\right) $. In the absence of harmonic trapping potential, the underlying two parameters of the Luttinger liquid model -- the sound velocity $c$ and the Luttinger exponent of long-wavelength correlation functions $K$ -- can be determined by using the Bethe ansatz technique \cite{Guan2012}. With harmonic traps, we use the local density approximation and assume that the system has a local sound velocity $c(x)=\sqrt{[\partial \mu /\partial n]n(x)/m}$ (where $\mu$ is the chemical potential) and Luttinger parameter $K(x)=\pi [\partial n/\partial \mu ]c(x)$, through their dependence on the local nonuniform density profile $n(x)$.

By linearizing the Luttinger liquid Hamiltonian, we derive the hydrodynamic equation of motion \cite{Liu2004}: 
\begin{equation}
\frac{\partial ^2\delta n\left( x,t\right) }{\partial t^2}=\frac \partial {\partial x}\left[ c\left( x\right) K\left( x\right) \frac \partial {\partial x}\left( \frac{c\left( x\right) }{K\left( x\right) }\delta n\left(x,t\right) \right) \right] .  
\label{hydro}
\end{equation}
The low-energy collective oscillations of density fluctuations $\delta n\left( x,t\right) =\delta n\left( x\right) \exp \left( i\omega _mt\right) $ with frequency $\omega _m$ may be classified by the number of nodes ($m$) in their eigenfunction $\delta n\left( x\right) $. We have solved these eigenfunctions by using a multi-series-expansion method \cite{Liu2008}, under the boundary condition that the current $j\left( x\right) $ must vanish identically at the Thomas-Fermi boundary $x=\pm x_{TF}$. The lowest two collective modes with $m=1,2$ are the dipole and breathing (compressional) modes, respectively, which can be excited separately by shifting the trap center or modulating the harmonic trapping frequency. The dipole mode is not affected by interactions according to Kohn's theorem, and has an invariant frequency precisely at the trap frequency $\omega_1=\omega_x$. Therefore, the frequency of the breathing mode $\omega _B=\omega _2$ provides the first mean to probe the non-trivial thermodynamics of our interacting Fermi system. In the case of a two-component Fermi gas, the breathing mode frequency has been calculated by Astrakharchik and colleagues by using a sum-rule approach \cite{astrakharchik2004}. 

With increasing number of spin components $N$, the constraints of the Pauli exclusion principle become less stringent, making the Fermi system to acquire a more ``bosonic'' behaviour. Indeed, it was shown very recently by Yang and You that in the limit $N\rightarrow \infty $ multi-component 1D fermions have the same energy per particle as spinless bosons \cite{yang2011}. We have calculated the breathing mode frequency of a 1D spinless Bose gas by solving the hydrodynamic equation Eq. (\ref{hydro}), with the local sound velocity and Luttinger parameter obtained from the Lieb-Liniger solution \cite{lieb1963}.


\begin{thebibliography}{99}

\bibitem[1]{giamarchi2004}
Giamarchi, T. {\it Quantum Physics in One Dimension} (Oxford University Press, 2004)

\bibitem[2]{yang1967}
Yang, C. N. Some exact results for the many-body problem in one dimension with repulsive delta-function interaction. {\it Phys. Rev. Lett.} {\bf 19}, 1312--1315 (1967).

\bibitem[3]{gaudin1967}
Gaudin, M. Un systeme a une dimension de fermions en interaction. {\it Phys. Lett. A} {\bf 24}, 55--56 (1967).

\bibitem[4]{sutherland1968}
Sutherland, B. Further results for the many-body problem in one dimension. {\it Phys. Rev. Lett.} {\bf 20}, 98--100 (1968).

\bibitem[5]{luttingerreview}
Voit, J. One-dimensional Fermi liquids. {\it Rep. Prog. Phys.} {\bf 58}, 977--1116 (1995).

\bibitem[6]{fiete2007}
Fiete, G. A. The spin-incoherent Luttinger liquid. {\it Rev. Mod. Phys.} {\bf 79}, 801--820 (2007).

\bibitem[7]{imambekov2012}
Imambekov, A., Schmidt, T. L. \& Glazman, L. I., One-dimensional quantum liquids: beyond the Luttinger liquid paradigm. {\it Rev. Mod. Phys.} {\bf 84}, 1253--1306 (2012). 

\bibitem[8]{recati2003}
Recati, A., Fedichev, P. O., Zwerger, W. \& Zoller, P. Spin-charge separation in ultracold quantum gases. {\it Phys. Rev. Lett.} {\bf 90}, 020401 (2003).

\bibitem[9]{lieb1963}
Lieb, E. H. \& Liniger, W. Exact analysis of an interacting Bose gas. I. The general solution and the ground state. {\it Phys. Rev.} {\bf 130}, 1605--1616 (1963). 

\bibitem[10]{girardeau1960}
Girardeau, M. Relationship between systems of impenetrable bosons and fermions in one dimension. {\it J. Math. Phys.} {\bf 1}, 516--523 (1960).

\bibitem[11]{mceuen1999}
Bockrath, M., Cobden, D. H., Lu, J., Rinzler, A. G., Smalley, R. E., Balents, L. \& McEuen, P. L. Luttinger-liquid behaviour in carbon nanotubes. {\it Nature} {\bf 397}, 598--601 (1999). 

\bibitem[12]{west1996}
Yacoby, A., Stormer, H. L., Wingreen, N. S., Pfeiffer, L. N., Baldwin, K. W. \& West, K. W. Nonuniversal conductance quantization in quantum wires. {\it Phys. Rev. Lett.} {\bf 77}, 4612--4615 (1996). 

\bibitem[13]{paredes2004}
Paredes, B., Widera, A., Murg, V., Mandel, O., F{\"o}lling, S., Cirac, I., Shlyapnikov, G. V., H{\"a}nsch, T. W. \& Bloch, I. Tonks--Girardeau gas of ultracold atoms in an optical lattice. {\it Nature} {\bf 429}, 277--281 (2004).

\bibitem[14]{weiss2004}
Kinoshita, T., Wenger, T. \& Weiss, D. S. Observation of a one-dimensional Tonks-Girardeau gas. {\it Science} {\bf 305}, 1125--1128 (2004).

\bibitem[15]{ho2013}
Cui, X. \& Ho, T. L. Ground-state ferromagnetic transition in strongly repulsive one-dimensional Fermi gases. Preprint at http://arxiv.org/abs/1305.6361 (2013). 

\bibitem[16]{moritz2005}
Moritz, H., St\"oferle, T., G\"unter, K., K\"ohl, M. \& Esslinger, T. Confinement induced molecules in a 1D Fermi gas. {\it Phys. Rev. Lett.} {\bf 94}, 210401 (2005).

\bibitem[17]{hulet2010}
Liao, Y.-A., Rittner, A. S. C., Paprotta, T., Li, W., Partridge, G. B., Hulet, R. G., Baur, S. K. \& Mueller, E. J., Spin-imbalance in a one-dimensional Fermi gas. {\it Nature} {\bf 467}, 567--569 (2010).

\bibitem[18]{jochim2012}
Z{\"u}rn, G., Serwane, F., Lompe, T., Wenz, A. N., Ries, M. G., Bohn, J. E. \& Jochim, S. Fermionization of two distinguishable fermions. {\it Phys. Rev. Lett.} {\bf 108}, 075303 (2012).

\bibitem[19]{cazalilla2009}
Cazalilla, M. A., Ho, A. F. \& Ueda, M. Ultracold gases of ytterbium: ferromagnetism and Mott states in an SU(6) Fermi system. {\it New J. Phys.} {\bf 11}, 103033 (2009).

\bibitem[20]{gorshkov2010}
Gorshkov, A. V., Hermele, M., Gurarie, V., Xu, C., Julienne, P. S., Ye, J., Zoller, P., Demler, E., Lukin, M. D. \& Rey, A. M. Two-orbital SU(N) magnetism with ultracold alkaline-earth atoms. {\it Nat. Phys.} {\bf 6}, 289--295 (2010).

\bibitem[21]{taie2012}
Taie, S., Yamazaki, R., Sugawa, S. \& Takahashi, Y. An SU(6) Mott insulator of an atomic Fermi gas realized by large-spin Pomeranchuk cooling. {\it Nat. Phys.} {\bf 8}, 825--830 (2012).

\bibitem[22]{troyer1999}
Frischmuth, B., Mila, F. \& Troyer, M. Thermodynamics of the one-dimensional SU(4) symmetric spin-orbital model. {\it Phys. Rev. Lett.} {\bf 82}, 835--838 (1999). 

\bibitem[23]{bonnes2012}
Bonnes, L., Hazzard, K. R. A., Manmana, S. R., Rey, A. M. \& Wessel, S. Adiabatic loading of one-dimensional SU(N) alkaline-earth-atom fermions in optical lattices. {\it Phys. Rev. Lett.} {\bf 109}, 205305 (2012).

\bibitem[24]{messio2012}
Messio, L. \& Mila, F. Entropy dependence of correlations in one-dimensional SU(N) antiferromagnets. {\it Phys. Rev. Lett.} {\bf 109}, 205306 (2012).

\bibitem[25]{banerjee2013}
Banerjee, D., B\"ogli, M., Dalmonte, M., Rico, E., Stebler, P., Wiese, U.-J. \& Zoller, P. Atomic quantum simulation of U(N) and SU(N) non-Abelian lattice gauge theories. {\it Phys. Rev. Lett.} {\bf 110}, 125303 (2013).

\bibitem[26]{ogata1990}
Ogata, M. \& Shiba, H. Bethe-ansatz wave function, momentum distribution, and spin correlation in the one-dimensional strongly correlated Hubbard model. {\it Phys. Rev. B} {\bf 41}, 2326--2338 (1990).

\bibitem[27]{cheianov2005}
Cheianov, V. V., Smith, H. \& Zvonarev, M. B. Low-temperature crossover in the momentum distribution of cold atomic gases in one dimension. {\it Phys. Rev. A} {\bf 71}, 033610 (2005).

\bibitem[28]{feiguin2010}
Feiguin, A. E. \& Fiete, G. A. Spectral properties of a spin-incoherent Luttinger liquid. {\it Phys. Rev. B} {\bf 81}, 075108 (2010). 

\bibitem[29]{astrakharchik2004}
Astrakharchik, G. E., Blume, D., Giorgini, S. \&  Pitaevskii, L. P. Interacting fermions in highly elongated harmonic traps. {\it Phys. Rev. Lett.} {\bf 93}, 050402 (2004).

\bibitem[30]{yang2011}
Yang, C. N. \& You, Y.-Z. One-dimensional $w$-component fermions and bosons with repulsive delta function interaction. {\it Chin. Phys. Lett.} {\bf 28}, 020503 (2011).

\end{thebibliography}

\begin{thebibliography}{99}

\bibitem[31]{taie2010}
Taie, S., Takasu, Y., Sugawa, S., Yamazaki, R., Tsujimoto, T., Murakami, R. \& Takahashi, Y.  Realization of a SU(2)$\times$SU(6) system of fermions in a cold atomic gas. {\it Phys. Rev. Lett.} {\bf 105}, 190401 (2010).

\bibitem[32]{olshanii1998}
Olshanii, M. Atomic scattering in the presence of an external confinement and a gas of impenetrable bosons. {\it Phys. Rev. Lett.} {\bf 81}, 938--941 (1998).

\bibitem[33]{recati2003b}
Recati, A., Fedichev, P. O., Zwerger, W. \& Zoller, P. Fermi one-dimensional quantum gas: Luttinger liquid approach and spin-charge separation. {\it J. Opt. B: Quantum Semiclass. Opt.} {\bf 5}, S55 (2003).

\bibitem[34]{demarco2001}
DeMarco, B., {\it Quantum Behavior of an Atomic Fermi Gas} (PhD Thesis, University of Colorado, 2001).

\bibitem[35]{stenger1999}
Stenger, J., Inouye, S., Chikkatur, A. P., Stamper-Kurn, D. M., Pritchard, D. E. \& Ketterle, W. Bragg spectroscopy of a Bose-Einstein condensate. {\it Phys. Rev. Lett.} {\bf 82}, 4569--4573 (1999).

\bibitem[36]{steinhauer2002}
Steinhauer, J., Ozeri, R., Katz, N. \& Davidson, N. Excitation spectrum of a Bose-Einstein condensate. {\it Phys. Rev. Lett.} {\bf 88}, 120407 (2002).

\bibitem[37]{clement2009}
Cl\'ement, D., Fabbri, N., Fallani, L., Fort, C. \& Inguscio, M. Exploring correlated 1D Bose gases from the superfluid to the Mott-insulator state by inelastic light scattering. {\it Phys. Rev. Lett.} {\bf 102}, 155301 (2009).

\bibitem[38]{brunello2001}
Brunello, A., Dalfovo, F., Pitaevskii, L., Stringari, S. \& Zambelli, F. Momentum transferred to a trapped Bose-Einstein condensate by stimulated light scattering. {\it Phys. Rev. A} {\bf 64}, 063614 (2001).

\bibitem[39]{hoinka2012}
Hoinka, S., Lingham, M., Delehaye, M. \& Vale, C. J. Dynamic spin response of a strongly interacting Fermi gas. {\it Phys. Rev. Lett.} {\bf 109}, 050403 (2012).

\bibitem[40]{Guan2012}  
Guan, X.-W., Ma, Z.-Q. \& Wilson, B. One-dimensional multicomponent fermions with $\delta $-function interaction in strong- and weak-coupling limits: $\kappa $-component Fermi gas. {\it Phys. Rev. A} {\bf 85}, 033633 (2012).

\bibitem[41]{Liu2004}  
Liu, X.-J., Drummond, P. D. \& Hu, H. Signature of Mott-insulator transition with ultracold fermions in a one-dimensional optical lattice. {\it Phys. Rev. Lett}. {\bf 94}, 136406 (2005).

\bibitem[42]{Liu2008}  
Liu, X.-J., Hu H. \& Drummond, P. D. Multi-component strongly attractive Fermi gas: a color superconductor in a one-dimensional harmonic trap. {\it Phys. Rev. A} {\bf 77}, 013622 (2008).


\end{thebibliography}
\end{document}